\documentclass[prd,onecolumn,aps,superscriptaddress,nofootinbib,hidepacs]{revtex4}
\usepackage[]{epsfig}
\usepackage[centertags]{amsmath}
\usepackage{amsfonts}
\usepackage{amssymb}
\usepackage[english]{babel}

\newcommand{\be}{\begin{equation}}
\newcommand{\ee}{\end{equation}}
\newcommand{\bea}{\begin{eqnarray}}
\newcommand{\eea}{\end{eqnarray}}
\newcommand{\ba}{\begin{eqnarray}}
\newcommand{\ea}{\end{eqnarray}}

\begin{document}
%%%%%%%%%%%%%%%%%%%%%%%%%%%%%%%%%%%%%%%%%%%%%%%%%%%%%%%%%%%%%%%%%%%

\title{Local and Semi-local Vortices in Yang-Mills-Chern-Simons
model}

%%%%%%%%%%%%%%%%%%%%%%%%%%%%%%%%%%%%%%%%%%%%%%%%%%%%%%%%%%%%%%%%%%%%%%%%%%%%%%%
\author{M.~Buck}
\affiliation{Departamento de F\'{\i}sica, Facultad de Ciencias
Exactas\\
Universidad Nacional de La Plata, C.C 67, 1900 La Plata,
Argentina}
\affiliation{Departamento de F\'\i sica, FCEyN, Universidad de
Buenos Aires\\
Pab.1, Ciudad Universitaria, Buenos Aires, Argentina}
%%%%%%%%%%%%%%%%%%%%%%%%%%%%%%%%%%%%%%%%%%%%%%%%%%%%%%%%%%%%%%%%%%%%%%%%%%%%%%%
\author{E.~F.~Moreno}
\affiliation{Department of Physics, West Virginia
University\\
Morgantown, West Virginia 26506-6315, U.S.A.}
%%%%%%%%%%%%%%%%%%%%%%%%%%%%%%%%%%%%%%%%%%%%%%%%%%%%%%%%%%%%%%%%%%%%%%%%%%%%%%%
\author{F.~A.~Schaposnik}
\thanks{Associated with CICBA}
\affiliation{Departamento de F\'{\i}sica, Facultad de Ciencias
Exactas\\
Universidad Nacional de La Plata, C.C 67, 1900 La Plata,
Argentina}
%

%%%%%%%%%%%%%%%%%%%%%%%%%%%%%%%%%%%%%%%%%%%%%%%%%%%%%%%%%%%%%%%%%%%
\begin{abstract}
We study BPS vortex configurations in three dimensional $U(N)$
Yang-Mills theories with Chern-Simons interaction coupled to
scalar fields carrying flavor. We consider two kind of
configurations: local vortices (when the number of flavors
$N_f=N$), and semi-local vortices (when $N_f>N$). In both cases
we carefully analyze the electric and magnetic properties and
present explicit numerical solutions.
\end{abstract}


 \maketitle
%%%%%%%%%%%%%%%%%%%%%%%%%%%%%%%%%%%%%%%%%%%%%%%%%%%%%%%%%%%%%%%%%%%

\section{Introduction}
Non-Abelian vortices may find their application in a variety of
problems ranging from particle physics and cosmology (e.g.
confinement, supersymmetric and supergravity models, hot or
dense QCD, cosmic strings) to condensed matter physics (e.g.
quantum Hall effect). Of particular interest are those vortices
solving first order BPS equations, which correspond to the
saturation of the Bogomolny bound for the mass and naturally
arise in supersymmetric theories (see refs.\cite{T}-(\cite{SY}
for reviews with complete lists of references).

Non-Abelian BPS equations for vortices  have been analyzed
both for Yang-Mills-Higgs \cite{Hanany}-\cite{Gorsky} and  for
Chern-Simons-Higgs \cite{deVS1}-\cite{Navarro} models. In the
former case   the gauge and the (fourth order) Higgs potential
coupling constants have to be related in order to pass from the
second order equations of motion to a first order BPS system.
When the Chern-Simons term dictates the dynamics of the gauge
field, one is forced to choose a sixth order Higgs potential in
order to find a Bogomolny bound for the vortex mass. Again,
coupling constants should be related
\cite{HKP}-\cite{JW}, \cite{CLMS}.

The origin of  these requirements can be also
understood in the framework of supersymmetry: they are
necessary conditions for the existence of an ${\cal  N}=2$
supersymmetry extension of the bosonic models. In this context,
the first-order BPS equations arise studying the supersymmetry
algebra and looking for supersymmetric states. The resulting
selfdual and anti-selfdual solutions break $1/2$ of the
original supersymmetry \cite{Witten}-\cite{ENS}.

The mixed case of Yang-Mills-Chern-Simons (YMCSS) vortices  was
also recently discussed \cite{CollieTong}-\cite{Collie}. As in
the abelian case \cite{LeeLeeMin}, in order to have a Bogomolny
bound and first order BPS equations the coexistence of the two
terms giving dynamics to the gauge field requires a careful
choice of the number and type of scalars. Moreover, the Gauss
law, through which the Chern-Simons term enters into the
energy, is no more an algebraic equation for $A_0$, as in the
case when the Yang-Mills term is absent,  but a second order
differential equation that should be taken into account
together with the first order BPS system.

It is the purpose of this work to construct explicit  BPS
vortex  solutions for the the YMCSS model thus completing the
analysis presented in \cite{Collie} where the low-energy vortex
dynamics was the main aim of study. We shall  take $U(N)$ as
gauge group and include $N_f$ scalars in the fundamental
representation  and one real scalar in the adjoint
representation of   $U(N)$. We shall consider both cases: when
the number of  flavors  $N_f$ is equal to the number of colors
$N$ and also when $N_f> N$, in which case vortices become
semi-local \cite{SY2},\cite{L},\cite{E1}-\cite{Auzzi2}.

\section{The model and the BPS equations}
We shall consider the $d=2+1$ dimensional U(N)
Yang-Mills-Chern-Simons model discussed in \cite{Collie} with
dynamics governed by the Lagrangian
\begin{equation}
\begin{split}
\mathcal{L} = &-\frac{1}{2e^2}
\text{Tr}F_{\mu\nu}F^{\mu\nu}-\frac{\kappa}{4\pi}
\text{Tr}\epsilon^{\mu\nu\rho}\left(A_\mu \partial_\nu A_\rho -
\frac{2i}{3}A_\mu A_\nu A_\rho\right)+\frac{1}{e^2}\text{Tr}(\mathcal{D}_\mu\phi)^2  \\
& +\left|\mathcal{D}_\mu q_i\right|^2 - q_i^{\dagger}(\phi-m_i)^2q_i -
\frac{e^2}{4}\text{Tr}\left(q_i q_i^{\dagger} - \frac{\kappa\phi}{2\pi}-v^2\right)^2.
\end{split}
\label{lagrangian}
\end{equation}
Here  $q_i$ are $N_f$ scalars   with $i$ the flavor index ($i =
1,2,\ldots, N_f$). Each $q_i$ transforms in the fundamental
representation of the gauge group $U(N)$ and  $\phi$ is a real
scalar in the adjoint. We shall first discuss the $N_f=N$ case
(local vortices) and then extend the analysis to $N_f> N$
(semilocal vortices) Whenever it does not lead to confusion
summation over flavors is implicit. The gauge field $A_\mu$
takes values in the Lie algebra of $U(N),~ A_\mu = A_\mu^A
t^A$, where $t^A =\left(t^A\right)^{ab}$ are the $U(N)$
generators ($A=1,\ldots,N^2-1; ~a,b=1,\ldots N$) { with
normalization} $\text{Tr}\,t^At^B=\delta^{AB}/2$. The curvature
and covariant derivatives are defined as
\begin{align}
F_{\mu\nu} &= \partial_\mu A_\nu - \partial_\nu A_\mu - i
\left[A_\mu, A_\nu \right] \nonumber\\
\mathcal{D}_{\mu}\phi &= \partial_\mu\phi -i\left[A_\mu,\phi\right]
\nonumber\\
\mathcal{D}_{\mu} q_i &=  \partial_\mu q_i -iA_\mu q_i
\label{deri}
\end{align}
The Chern-Simons coefficient  $\kappa$ must be an integer for $N>1$.

The masses  $ {m_i }$ of the fundamental scalars break the
flavor symmetry to $U(1)^{N-1}_f.$ There exists a fully broken
Higgs phase in which the scalars take the following expectation
values
\begin{equation}
q_{ i\, \text{vac}}^a=\delta_i^a
\sqrt{v^2+\frac{\kappa m_i}{2\pi}} \qquad  \phi_{\text{vac}}^{ab}
=\delta^{ab}m_b \qquad  A_{\mu \, \text{vac} }^{ab} = 0. \label{qvac}
\end{equation}
In this vacuum, the $U(N)$ gauge symmetry and the
$U(1)^{N-1}_f$ flavour symmetry are spontaneously broken. There
remains only a diagonal symmetry, $U(N) \times U(1)^{N-1}_f
\rightarrow U(1)^{N-1}_{\text{diag}}$. This corresponds to the
combined action of a gauge group element $U^{ab}\in U(N)$ and a
flavour transformation $V_{ij}=\delta_{ij}e^{i\alpha_j}\in
U(1)^{N-1}_f$ (with $N-1$ independent parameters $\alpha_j$):
\begin{equation}
q_{i\,\text{vac}}^a \rightarrow U^{ab}  q_{j\, \text{vac}} ^b V_{ji} \qquad\qquad
\phi_{\text{vac}}^{ab} \rightarrow U^{ac}  \phi_{\text{vac}} ^{cd}\left(U^{-1}\right)^{db}
\end{equation}
where $U=V^{-1}$.

Let us note that in the $e^2 \to \infty$ limit where the
Yang-Mills term and the kinetic energy term for the $\phi$
field can be discarded, the adjoint field $\phi$ may be
eliminated from the Lagrangian, and for $m_i = 0$ one ends up
with the sixth order potential which allows one to find BPS
equations for the pure CS-Higgs system both in the Abelian
\cite{HKP}-\cite{JW} and non-Abelian \cite{CLMS} cases,
\begin{equation}
\lim_{e^2 \to \infty} V[q,\phi, m_i=0] = \frac{(4\pi)^2}{\kappa^2}
\left(\left|q_i\right|^2 - v^2\right)^2 \left|q_i\right|^2.
\end{equation}

The energy  associated with Lagrangian (\ref{lagrangian}) can
be constructed from $T_{00}$,  the time-time component of the energy
momentum tensor $T_{\mu\nu}$. It takes the form
\begin{equation}
\begin{split}
E =\int\,d^2x\, T_{00} = \int\,d^2x\,& \left[
\frac{1}{e^2}\text{Tr}\left(E_\alpha^2+B^2\right) +
\frac{1}{e^2}\text{Tr}\left((\mathcal{D}_0\phi)^2+(\mathcal{D}_\alpha\phi)^2\right)
+\left|\mathcal{D}_0 q_i\right|^2 +
\left|\mathcal{D}_\alpha q_i\right|^2\right.\label{boundi} \\
&+ \left.q_i^{\dagger}(\phi-m_i)^2q_i +\frac{e^2}{4}
\text{Tr}\left(q_i q_i^{\dagger} -
\frac{\kappa\phi}{2\pi}-v^2\right)^2 \right]
\end{split}
\end{equation}
where $E_\alpha=F_{0\alpha}$, $B=F_{12}$.

Following the Bogomolny procedure of square completion a lower
energy bound can be
 obtained \cite{Collie}
\begin{equation}
E \geq \left\vert
2\pi n v^2 + \sum_i Q_im_i.
\right\vert
\label{bound}
\end{equation}
Here $n\in\mathbb{Z}$ is the topological charge of the
configuration, which corresponds to the topological degree
associated to the $q_i$ component that carries the winding,
\begin{equation}
n = \frac{1}{2\pi} {\rm Tr}\int d^2x B
\end{equation}
and $Q_i$ are the conserved Noether charges associated with the
residual $U(1)^{N-1}$ flavor symmetry,
\begin{equation}
Q_i = i \int d^2x \left (q_i^\dagger {\cal D}_0 q_i -
\left({\cal D}_0 q_i\right)^\dagger q_i
\right).
\end{equation}
Using Gauss law one can find the typical Chern-Simons term
connection between charge and flux,
\be \sum_i Q_i =
\frac{e\kappa}{2\pi} \Phi \label{qes} \ee with $\Phi$ the magnetic
flux, \be \Phi = \frac1e {\rm Tr} \int d^ 2x B = \frac{2\pi}{e} n
\ee

The bound (\ref{bound}) is saturated whenever the following BPS
first-order equations hold
\begin{align}
\mathcal{D}_1q_i\pm i\mathcal{D}_2q_i &=0 \label{b1}\\
B \pm \frac{e^2}{2}(q_iq_i^\dagger-\frac{\kappa\phi}{2\pi}-v^2) &=0 \label{b2}\\
\mathcal{D}_0\phi &=0 \label{b3}\\
E_\alpha\mp \mathcal{D}_\alpha\phi &= 0 \label{b4}\\
\mathcal{D}_0q_i\mp i(\phi-m_i)q_i &=0. \label{b5}
\end{align}
It should be signaled that in obtaining the bound, it is
necessary to use Gauss' law~\cite{Collie}
\begin{equation}
-\frac{\kappa}{4\pi}B+\frac{i}{2}\left[\left(\mathcal{D}_0 q_i\right)
q_i^\dagger-q_i\left(\mathcal{D}_0q_i\right)^\dagger\right]+
\frac{1}{e^2}\mathcal{D}_\alpha E_\alpha + \frac{i}{e^2}
\left[\mathcal{D}_0\phi,\phi\right]=0.
\label{gauss}
\end{equation}
which should then be considered together with
eqs.(\ref{b1})-(\ref{b5}) when looking for explicit vortex
solutions.

At the bound, the energy of configurations can be identified
with the BPS soliton mass,
\begin{equation}
M=\left|
2\pi v^2 n \,\,+ \,\, \sum_i Q_im_i \right|\label{boundsat}
\end{equation}
In what follows we choose  the upper sign in
eqs.(\ref{b1})-(\ref{b5}), which corresponds to a non-negative
winding. Of course, the opposite choice is equally treatable.

\section{The vortex ansatz}

Starting from the trivial vacuum, a winding can be introduced
through a singular gauge transformation generated by
\begin{equation}
\Omega(\varphi)=\text{diag}\left[1,1,...,e^{in\varphi}\right] =
e^{\frac{in\varphi}{N}}
\text{diag}[e^{-i\frac{n}{N}\varphi},e^{-i\frac{n}{N}\varphi}
,...,e^{i\frac{n(N-1)}{N}\varphi}].
\end{equation}
We have written the formula above, so as to emphasize that $\Omega$  combines an
$U(1)$ element with an element of $\mathbb{Z}_N$, the center of
$SU(N)$. Then, a configuration of the form
\begin{equation}
q_{sing}  =   \Omega(\varphi)  q_{vac}
\end{equation}
with $q_{vac}$ the trivial vacuum (\ref{qvac}) will lead to a
topologically nontrivial but singular (at the origin) string
configuration. To avoid the singularity  the natural ansatz for
a regular vortex should be
\begin{equation}
q =\text{diag}\left[\eta_1,\eta_2,...,\eta_N e^{in\varphi} q_N(\rho)\right]
\label{ansatz1}
\end{equation}
with with $\eta_i^2\equiv v^2+ {\kappa m_i}/{2\pi}$ and
$q_N(\rho)$ vanishing at $\rho = 0$. Then, (\ref{ansatz1})
should be supplemented with consistent ansatz for the remaining
fields,
\begin{eqnarray}
\phi &=&\text{diag}\left[m_1,m_2,...,m_N+h_N(\rho)\right] \\
A_\varphi &=&\text{diag}\left[0,0,...,n-a_N(\rho)\right] \\
A_\rho &=&0,
\label{ansatz2}
\end{eqnarray}
The complete set of appropriate boundary conditions ensuring
finite energy is:
\begin{eqnarray}
a_N(0) &=& n  \; , \;\;\;\; a_N(\infty)= 0 \; , \;\;\;\;
q_N(0) = 0 \\
q_N(\infty) &=& 1  \; , \;\;\;\;
h_N(\infty) = 0.
\end{eqnarray}
Concerning   $A_0$,   equations (\ref{b3}) and (\ref{b4}) require
\begin{equation}
\left[A_0,\phi\right]=0 \qquad \partial_\rho\left(A_0+\phi\right)=0.
\end{equation}
This suggests
\begin{equation}
A_0=-\phi+C \quad \text{with} \quad \left[C,\phi\right]=0,
\end{equation}
with $C$   determined by (\ref{b5}):
\begin{equation}
\left(\phi+A_0\right)^{ab}q_i^b=m_iq_i^a \quad \rightarrow \quad
C^{ab}=\delta^{ab}m_b.\label{C}
\end{equation}
Unless all masses vanish, one cannot set $A_0=-\phi$.  The
above equation fixes $A_0$ in our ansatz:
\begin{equation}
A_0 =\text{diag}\left[0,0,...,-h_N(\rho)\right].
\end{equation}

\section{The BPS vortex solution}
Plugging   ansatz ~(\ref{ansatz1})-(\ref{ansatz2}) into
equations (\ref{b1}), (\ref{b2}) and (\ref{gauss}) gives
\begin{align}
\rho\,\partial_\rho q_N - a_Nq_N &= 0 \label{e1}\\
\frac{1}{\rho}\,\partial_\rho a_N-\frac{e^2}{2}\left(\eta_N^2q_N^2 -
\frac{\kappa}{2\pi}h_N-\eta_N^2\right) &= 0\label{e2}\\
\frac{\kappa}{4\pi\rho}\,\partial_\rho a_N
-h_N\eta_N^2q_N^2 + \frac{1}{e^2}\left(\partial_\rho^2 h_N +
\frac1\rho \,\partial_\rho h_N
\right )  &= 0\label{e3},
\end{align}
Note that the Gauss Law constraint is at  the origin of the
second-order derivative in $h_N$ in (\ref{e3}). It will be
convenient to define
\begin{equation}
\beta = \frac{e^2 \eta_N^2}{2} \qquad \gamma =
\frac{\kappa}{2\pi\eta_N^2}\\
\end{equation}
and
\begin{equation}
a(\tau) = a_N(\rho) \qquad q(\tau) = q_N(\rho) \qquad h(\tau) =
\gamma h_N(\rho)\label{param}
\end{equation}
where $\tau=\sqrt{\beta}\rho$. One can use (\ref{e1}) to
eliminate $\partial_\rho a_N$ in (\ref{e3}) so that equations
can be recasted in the form:
\begin{align}
\frac{da}{d\tau} &= \tau\left(q^2 -  h - 1\right) \label{br1}\\
\frac{dq}{d\tau} &= \frac{1}{\tau}\,qa \\
\frac{dh}{d\tau} &= u \\
\frac{du}{d\tau} &= 2hq^2 - \alpha\left(q^2 -   h -
1\right)-\frac{1}{\tau}\,u,\label{br4}
\end{align}
where $\alpha=\beta\gamma^2$. The boundary conditions imposed
by finite energy read now
\begin{eqnarray}
a(0) &=& n \; , \;\;\;\; a(\infty) = 0
\nonumber\\
q(0) &=& 0 \; , \;\;\;\; q(\infty)  =  1  \nonumber\\
h(\infty) &=& 0.
\end{eqnarray}
{Concerning the behavior of $h$ at the origin, it should go to
a finite constant}.

Using (\ref{boundsat}), the vortex mass for our ansatz takes
the form
\begin{equation}
M=2\pi v^2n + Q_Nm_N=2\pi\eta_N^2n.
\label{41}
\end{equation}
The BPS vortex mass is solely determined by the topological
charge, the r\^ole of $\eta_N^2$ being just that of a scale.

Our result is of course consistent with the re-parametrized
form of the energy
\begin{equation}
E=
\frac{2\pi}{e^2}\int \tau d\tau \left(\frac2{\gamma^2}\left(h'\right)^2+
\frac{\beta}{\tau^2}\left(a'\right)^2
+\frac4{\gamma^2}q^2h^2+2\frac{\beta}{\tau^2}a^2q^2+
2\beta\left(q'\right)^2+
\beta\left(q^2-
h -1\right)^2\right),
\end{equation}
which after some algebra and integration by parts can be
written as follows:
\begin{equation}
E= \frac{2\pi}{e^2}\int\tau d\tau \,  \left(
2\beta\left[q'\mp \frac{aq}{\tau}\right]^2+
\beta\left[\frac{a'}{\tau}\mp\left(q^2-h-1\right)\right]^2-
\frac{2}{\gamma^2}h\left[\frac{1}{\tau}\,\left(\tau h'\right)' -
2hq^2\pm\alpha \frac{a'}{\tau}\right] \mp 2\beta \tau^{-1}a'\right).
\end{equation}
The upper sign corresponds to our non-negative winding ansatz.
The first three terms in the integral vanish as they are
readily identified with the Bogomolny equations, while the last
term gives the expected contribution to the vortex mass
$M=2\pi\eta_N^2n$.\\

Let us note that for $m_N=0$,  eqs.(\ref{br1})-(\ref{br4})
reduce to the abelian case discussed in~\cite{LeeLeeMin}. It is
indeed typically observed that the non-abelian
$\mathbb{Z}_N$-vortex equations of a model reduce to the
abelian equations when the coupling constants of the $U(1)$-
and the $SU(N)$-gauge groups are set equal (this choice has
been made implicitly here by working with the gauge group
$U(N)$). In the case $m_N \neq 0$, the only modification with
respect to the $m_N=0$ case arises through the parameter
$\alpha =\frac{e^2}{4\pi} \kappa^2 \left(2\pi v^2 + \kappa
m_N\right)^{-1}$. Hence we can obtain the profile functions of
any general $\left\{m_N\neq0, \kappa\right\}$ vortex from a
$\left\{m_N=0,\kappa'\right\}$ solution by setting
$\kappa'=\kappa/\sqrt{1+\kappa m_N/2\pi v^2}$. However, the
behavior of the physical observables depends upon the value of
$m_N$. If the Chern-Simons term is absent ($\kappa =0$) the
Gauss law is satisfied by $h=0$ {and our ansatz reduces to the
well-honored Abrikosov-Nielsen-Olesen vortex}
\cite{A}-\cite{NO}.  \\

\begin{figure}[h] \center
\includegraphics[width=0.58\textwidth]{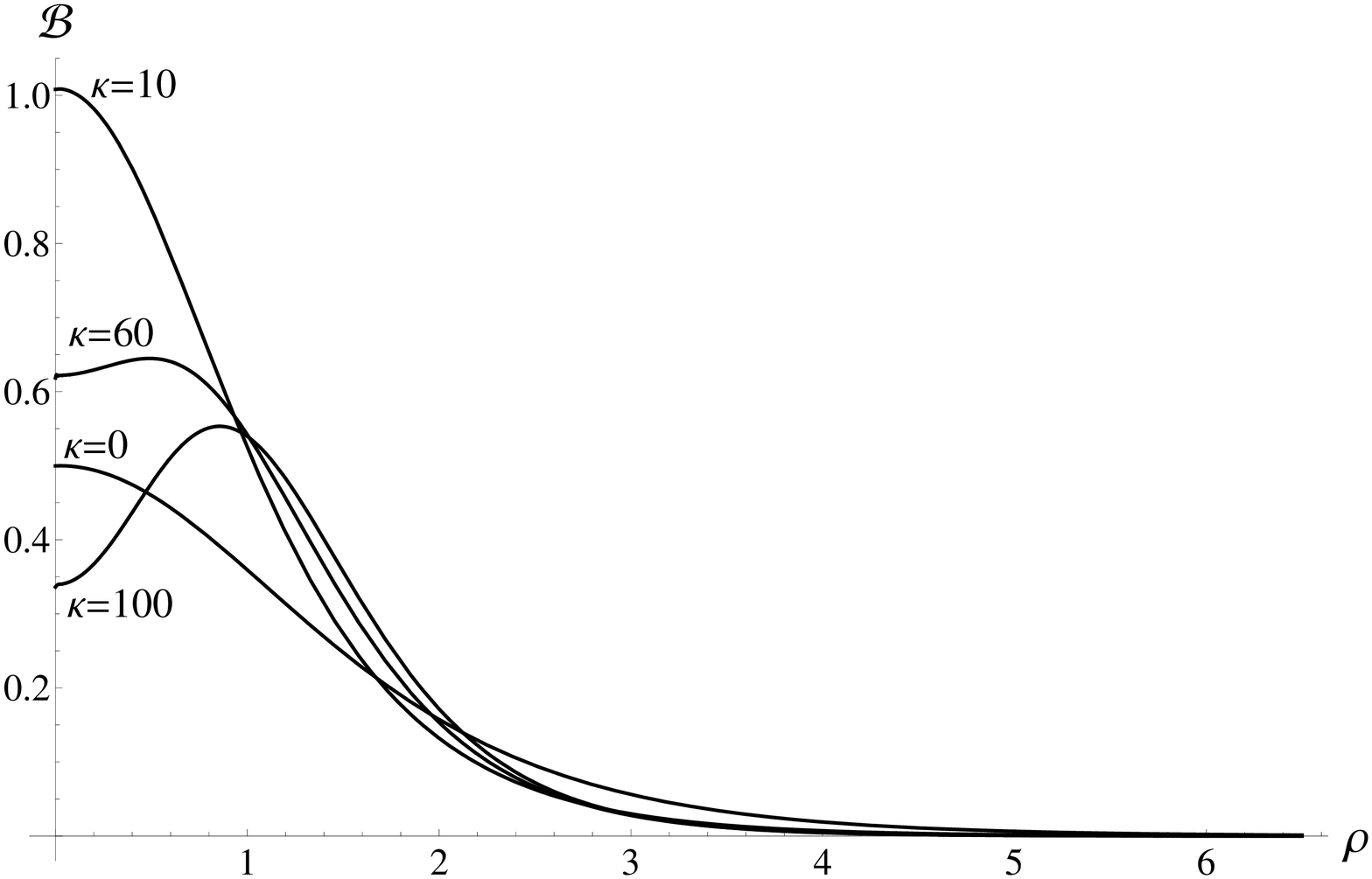}
\includegraphics[width=0.58\textwidth]{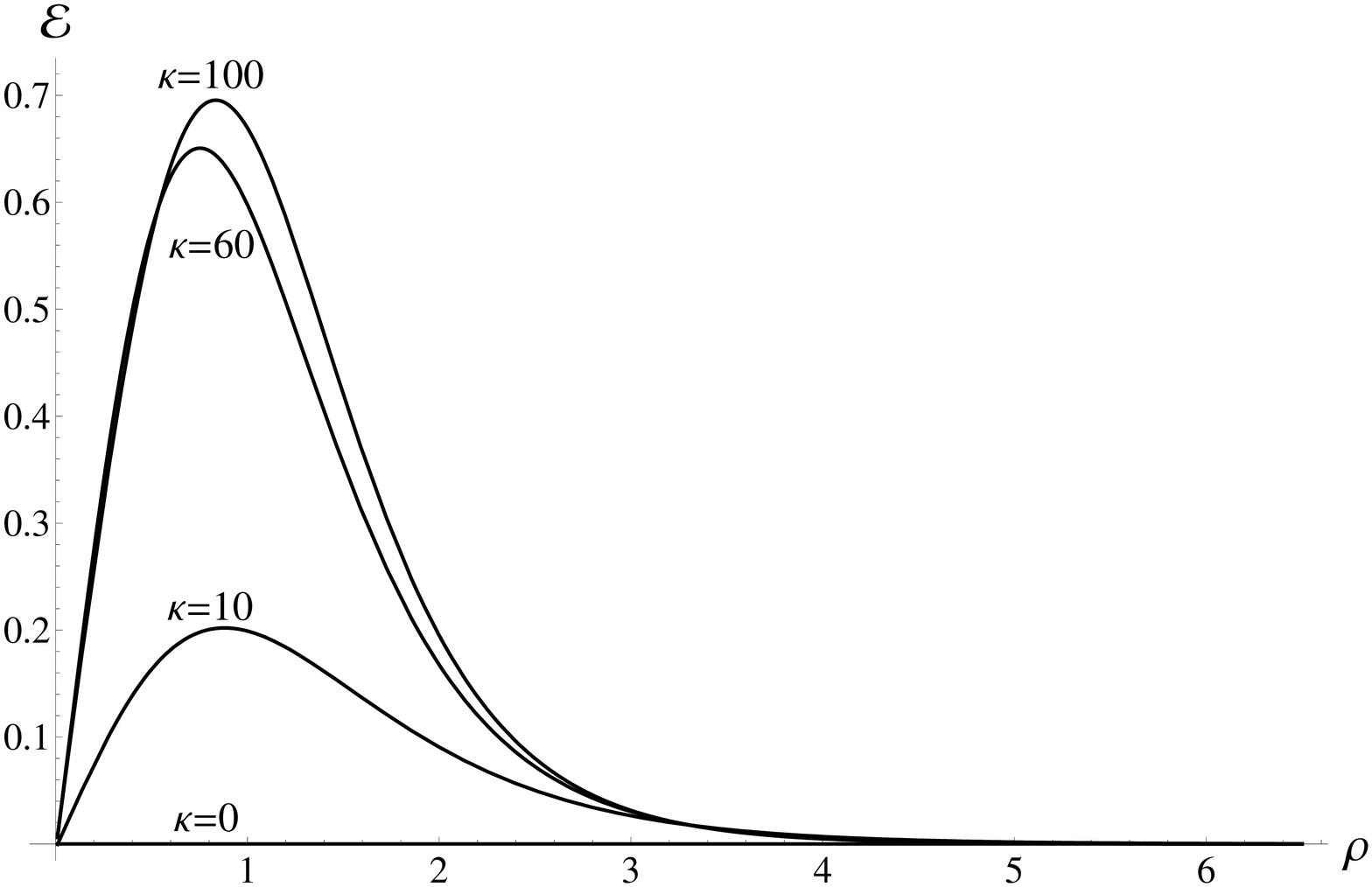}
\caption{The magnetic field {\cal B} and electric field {\cal E}
of the $\left\{n=1,m_N/v^2=1,\kappa\right\}$ vortex are shown for
$\kappa=0,10,60,100$. The $\kappa=0$ line corresponds to the
abelian Nielsen-Olesen vortex, which exhibits no electric field.}
\label{fig:bdlocal}
\end{figure}

{\noindent To obtain numerical solutions to the BPS equations
(\ref{br1})-(\ref{br4}) we have used a relaxation method,
selecting the following four boundary conditions
\begin{equation}
a(0) = 1, \quad q(\infty) = 1, \quad h(\infty) = u(\infty) = 0.
\end{equation}
Given our ansatz, the magnetic field ${\cal B}$ and the electric
field ${\cal E}$ can be defined as \be {\cal B}= {\rm Tr}F_{12} \; ,
\;\;\;\; {\cal E} = {\rm Tr}F_{0\rho}. \ee They are depicted in
Figure~\ref{fig:bdlocal} where we have set $e^2/v^2=1$.
}
It is interesting to observe the behaviour of the magnetic
field in the case $m_N\neq0$. As $\kappa$ is increased, the
magnitude of the magnetic field at the origin initially
increases. At large enough $\kappa$, however, the $B$-field
starts to decrease at the origin and begins to show a
characteristic \emph{bump}, as encountered in the Abelian
case~\cite{LeeLeeMin}. Any further increase in $\kappa$
amplifies the size of the bump. The electric field is
considerably smaller than the magnetic field for small
$\kappa$, but it also increases as the CS-coupling becomes
important. As in the $m_N=0$ case, the ratio
$\mathcal{E}_\text{max}(\kappa)/\mathcal{B}_\text{max}(\kappa)$
increases linearly with $\kappa$ for small $\kappa$, and
eventually tends to a constant.

We may explain this behavior by observing that we have
$\alpha(\kappa)\propto\kappa^2$ $\forall$ $\kappa$ when $m_N=0$
and $\alpha(\kappa)\propto\kappa$ for $\kappa\gg 2\pi v^2/m_N$
when $m_N\neq0$. The limit $\kappa\gg 2\pi v^2/m_N$ is where
the effects of non-zero $m_N$ become noticeable. The profile
functions $a,h$ and $q$ thus depend more sensitively on
$\kappa$ when the mass is zero. Furthermore, when computing the
electric and magnetic fields from the functions $a$ and $h$,
their behavior is not explicitly dependent on $\kappa$ for
$m_N=0$, whereas we have explicit dependence of order
$\mathcal{O}(\kappa)$ in $\mathcal{B}$ and of order
$\mathcal{O}(\sqrt{\kappa})$ in $\mathcal{E}$ for $m_N\neq0$ in
the same limit. In the zero-mass case~\cite{LeeLeeMin}, the
magnetic field starts to exhibit the typical doughnut-shape as
$\kappa$ is increased, at the same time as its overall strength
decreases. The electric field develops the characteristic bump
and also becomes weaker. As a consequence, the ratio
$\mathcal{E}_\text{max}/\mathcal{B}_\text{max}$, which vanishes
as $\kappa\rightarrow0$, approaches a finite constant for
$\kappa\gg 2\pi v^2/m_N$. For $m_N\neq0$, on the other hand,
the explicit $\kappa$-dependence of the electric and magnetic
fields counter-acts the trend of $a$ and $h$ becoming smaller
for larger $\kappa$. The maxima of the two fields now
increasing for a larger range of $\kappa$ and do not go to zero
as $\kappa\rightarrow\infty$. As a result, the ratio
$\mathcal{E}_\text{max}(\kappa)/\mathcal{B}_\text{max}(\kappa)$
still tends to a finite constant asymptotically, but it reaches
it more slowly as $m_N$ becomes larger. It is also worth noting
that the radius of the bump in the electric field linearly
increases with $\kappa$ when $m_N=0$, whereas it approaches a
constant when $m_N=0$ (see Figure~\ref{fig:bdlocal}).
\\

Let us end this section by computing the angular momentum for
the vortex solution using the formula
\begin{equation}
J = \int d^2x \, \varepsilon_{ij} x_i T_{0\,j} =  \int d^2x\,
T_{0\,\varphi}
\end{equation}
(which is actually independent of the ansatz). Given Lagrangian
(\ref{lagrangian}), one has
\begin{align}
T_{0 \varphi} &= \frac{2}{e^2}\rho \text{Tr}\left(E_\rho\, B\right) +
\frac{2}{e^2} \text{Tr}\left(D_0\phi\, D_\varphi
\phi\right) +
\left((D_0q_a)^\dagger\, (D_\varphi q_a) + (D_\varphi q_a)^\dagger\, (D_0
q_a)\right)
\nonumber\\
&=\frac{2}{e^2}\, h'_N\, a'_N + 2\, \eta_N^2\, h_N\, q_N^2\, a_N
\end{align}
Using Gauss' law, which for our ansatz reads
\be
\eta_N^2\, h_N\, q_n^2 =
\frac{1}{e^2}\left(h''_N + \frac{1}{\rho}\, h'_N\right)+
\frac{\kappa}{4\pi} (a'_N/\rho)
\ee
we can write
\begin{align}
T_{0 \varphi} &= \frac{2}{e^2\, \rho} \frac{d}{d\rho}\left( \rho\, a_N\,
h'_N\right)
+ \frac{\kappa}{4\pi}
\frac{1}{\rho} \frac{d (a_N^2)}{d\rho}
\end{align}
So, we have for $J$
\begin{equation}
J  = 2\pi \left.\left(\frac{2}{e^2}\rho\, a_N\, h'_N +
\frac{\kappa}{4\, \pi}\, a_N^2\right)\right|_0^\infty
\end{equation}
or, finally
\begin{equation} J =
-\frac{\kappa}{2}\, n^2
\end{equation}
Being $\kappa$ an integer, also the angular momentum is
quantized at the classical level. Note that in view of
eq.(\ref{qes}) which in the present case reduces to $ Q_N =
\kappa n $ we see that the angular momentum can be written in
terms of the square of the charge (in contrast with the pure
Yang-Mills case in which it is proportional to the charge
\cite{deVS1}-\cite{deVS2}).

\section{Semi-local Vortices}

Unlike Abrikosov-Nielsen-Olesen vortices and their non-abelian
extensions, the radius of semilocal vortex is not fixed but it
becomes a parameter. This kind of vortices emerge when $N_f>N$,
that is, when there are $N_e=N_f-N$ additional fundamental
scalars $\left\{p_e=q_{N+e}\right\}$, $e=1,...,N_f-N$ in
comparison with the local vortices arising in the $N=N_f$ case.

We then start from Lagrangian (\ref{lagrangian}) now with $N_f>
N$ and  consider minimization of the potential in the general
case $m_i \neq m_j$ $\forall$ $i\neq j$, with $i,j \leq N$.
The mass term in the $N_f >N$ Lagrangian
\begin{equation}
L_m = q_i^{\dagger a}\left(\phi^{ab}-\delta^{ab}m_i\right)^2q^b_i
\end{equation}
can only be made to vanish for $N$ non-zero fundamental
scalars, since we may only pick $N$ of the diagonal entries in
$\phi$ to cancel the $\delta^{ab}m_i$ terms. Then, in order to
minimize the potential the remaining $q_i$ need to vanish.
Without loss of generality, we may choose the
$\left\{p_e=q_{N+e}\right\}$ to vanish. These fields then lie
in the unbroken vacuum. Spontaneous symmetry breaking of
$U(N)_g\times U(1)^{N_f-1}_f$ occurs as before for the original
$N$-sector while  the additional fields exhibit invariance
under the more general transformation
\begin{equation}
p^a_{e\,\text{vac}} \longrightarrow \exp({i\alpha_e})U^{ab}
p^b_{e\,\text{vac}},
\end{equation}
with unconstrained $\left\{\alpha_e\right\}$. These fields must
be topologically trivial. If we adopt the previous ansatz for
the original fields, equation (\ref{b2}), which as we explain
below is still valid for $N_f> N$ requires all components of
$p^a_e$ to vanish identically except for $a=N$. This suggests
the following ansatz for the $p_e$
\begin{equation}
p^a_e=\eta_N\delta^{aN}\xi_e(\tau).
\label{augmented}
\end{equation}
{Furthermore, it is required by (\ref{b5}) that the masses of
any additional (non-trivial) scalar
$p_e$ be equal to the mass of the field that carries the winding,}
\begin{equation}
m_e=m_N \quad \text{if}\quad \xi_e(\tau)\neq 0.
\end{equation}

Equation (\ref{boundi}) for the energy in the $N_f = N$ case is
still valid for $N_f >N$ and so is the bound (\ref{bound}) and
the BPS equations (\ref{b1})-(\ref{b5}). Concerning the axially
symmetric ansatz, it consist of the one proposed in the local
case, eqs.(\ref{ansatz1})-(\ref{ansatz2}), augmented with
(\ref{augmented}) for the extra scalars. Inserting the ansatz
in the BPS equations one now obtains
\begin{align}
\frac{da}{d\tau} &= \tau\,(q^2 +\left|\xi_e\right|^2 -  h - 1)
\label{sys1} \\
\frac{dq}{d\tau} &= \frac{1}{\tau}\,qa \label{bp1}\\
\frac{d\xi_e}{d\tau} &= \frac{1}{\tau}\,(a-n)\xi_e
\; , \;\;\;\;\;\;  e=1,\ldots ,N_f-N_c\label{bp2}\\
\frac{dh}{d\tau} &= u \\
\frac{du}{d\tau} &= 2h(q^2+\left|\xi_e\right|^2) - \alpha
(q^2 +\left|\xi_e\right|^2 - h - 1) -\frac{1}{\tau}\,u,
\label{sysult}
\end{align}
with $\left|\xi_e\right|^2=\sum_e\xi_e^\dagger \xi_e$.
Equations (\ref{bp1}) and (\ref{bp2}) can be used
to solve for the profile functions~$\left\{\xi_e\right\}$ \cite{SY2}:
\begin{equation}
\xi_e(\tau)=\chi_e \frac{q(\tau)}{\tau^n}
\end{equation}
with $\chi_e\in\mathbb{C}$ arbitrary complex constants that
parametrize the solutions. Of course, if we set all the
$\chi_e$ parameters to zero the system
(\ref{sys1})-(\ref{sysult}) coincides with
(\ref{br1})-(\ref{br4})
and the semi-local vortices become ordinary local ones.\\

{\noindent The energy of the semi-local vortex is
\begin{align}
E_s=
\frac{2\pi}{e^2}\int \tau d\tau &\left(\frac2{\gamma^2}\left(h'\right)^2+
\frac{\beta}{\tau^2}\left(a'\right)^2
+\frac4{\gamma^2}h^2\left(q^2+\left|\xi_e\right|^2\right)+2\frac{\beta}{\tau^2}a^2q^2+
2\frac{\beta}{\tau^2}(a-n)^2\left|\xi_e\right|^2 \right.\nonumber\\
&\left. + 2\beta\left(\left(q'\right)^2+ \left|\xi'_e\right|^2\right) +
\beta\left(q^2+\left|\xi_e\right|^2 -
h -1\right)^2\vphantom{\frac4{\gamma^2}}\right).
\end{align}
As in the local case, it is straightforward to show that this
can be written as:
\begin{eqnarray}
E_s&=& \frac{2\pi}{e^2}\int\tau d\tau \,   \left(
2\beta\left[q'\mp \frac{aq}{\tau}\right]^2
+2\beta\left|\xi'_e\mp \frac{(a-n)\xi_e}{\tau}\right|^2
+\beta\left[\frac{a'}{\tau}\mp(q^2+\left|\xi_e\right|^2-h-1)\right]^2\right.\nonumber\\
&& \left.-\frac{2}{\gamma^2}h\left[\frac{1}{\tau}\,\left(\tau h'\right)' -
2h[q^2+\left|\xi_e\right|^2]\pm\alpha \frac{a'}{\tau}\right]
\mp 2\beta \tau^{-1}a'\vphantom{\left|\xi'_e\mp
\frac{(a-n)\xi_e}{\tau}\right|^2}\right),
\label{energys}
\end{eqnarray}
the upper sign corresponding to the non-negative winding
vortex. Energy (\ref{energys}) reduces to the lower bound
in~(\ref{bound}) when the Bogomolny equations are satisfied,
\begin{equation}
M_s  =  \left\vert
2\pi n v^2 + \sum_i Q_im_i
\right\vert
\label{boundsv}
\end{equation}

In the case of semilocal vortices, one can define a parameter
$\chi$, the complexified size of   the vortex through the
formula
\be
|\chi|^2 = \sum_e \left|\chi_e\right|^2
\ee
We thus see that, as expected,  the vortex mass is
$\chi$-independent but the behavior of the fields at infinity
drastically changes with respect to the local case: the fields
have a long range power falloff instead of an exponential one.
This can be seen in Figure~\ref{dbsemi}, which shows the
numerical solutions to (\ref{sys1})-(\ref{sysult}). Indeed, at
large distances ($\rho\gg 1/e\sqrt\beta$) and for very large
transverse size of the vortex ($\chi \gg e\sqrt\beta$) one can
see how the asymptotic behavior is no more that of an
exponential falloff but a power one. The analytical asymptotic
behavior of the fields $a_N$ and $q_N$ are
\begin{align}
a_n \approx n\, \frac{|\chi|^2}{\rho^{2|n|}}\; , \;\;\;\;
q_N \approx 1 - \frac{1}{2} \frac{|\chi|^2}{\rho^{2|n|}}
\end{align}
whereas $h_N$ has the same exponential falloff behavior as in
the local vortex case:
\be
h_N \approx \frac{e^{-\eta_N\, \rho}}{\sqrt{\rho}}
\ee

Let us finally note that the  presence of the radius $\chi$
also reduces the Chern-Simons characteristic bump at the origin.

\begin{figure}[h]
\center
\includegraphics[width=0.6\textwidth]{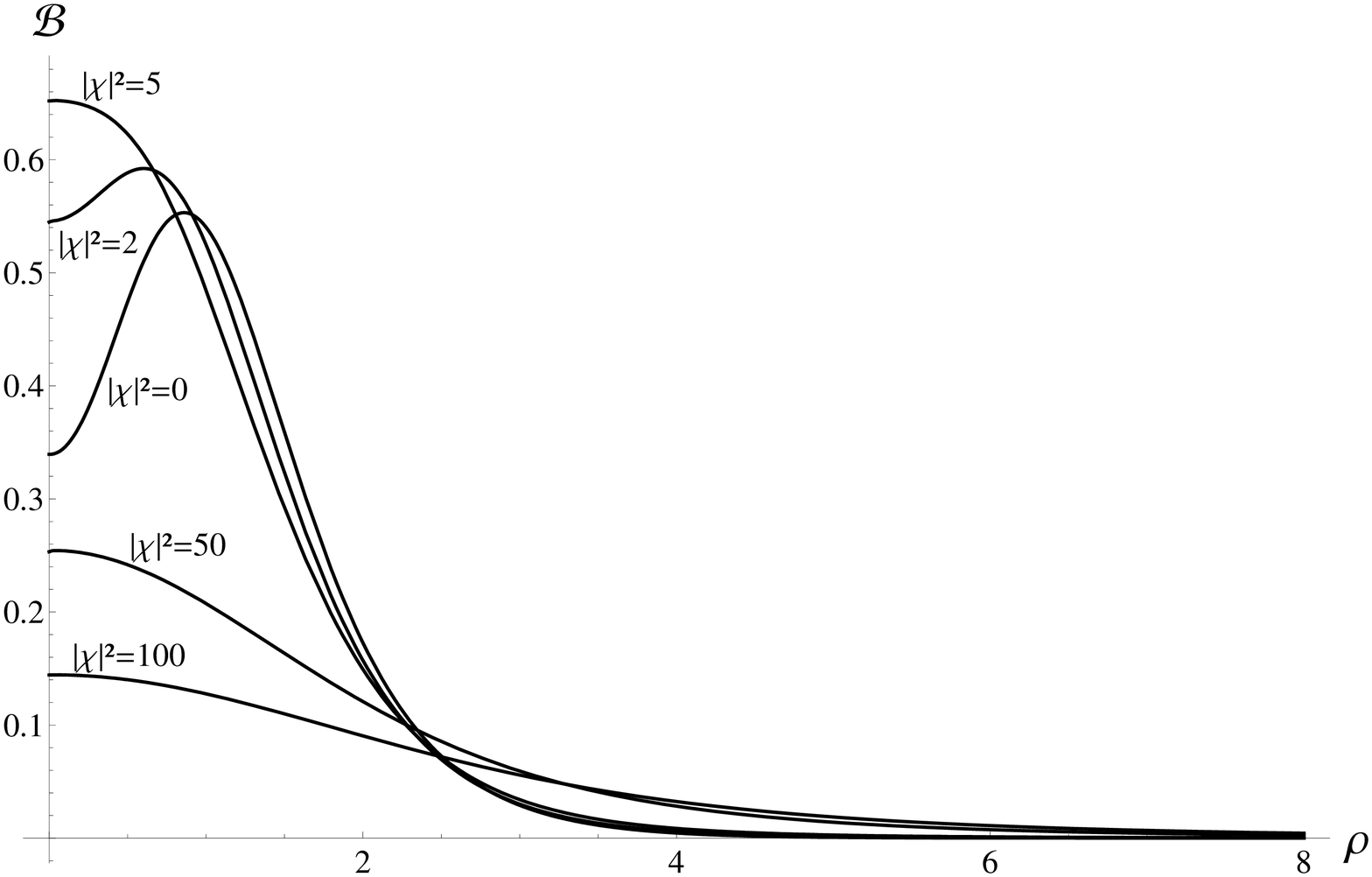}
\includegraphics[width=0.6\textwidth]{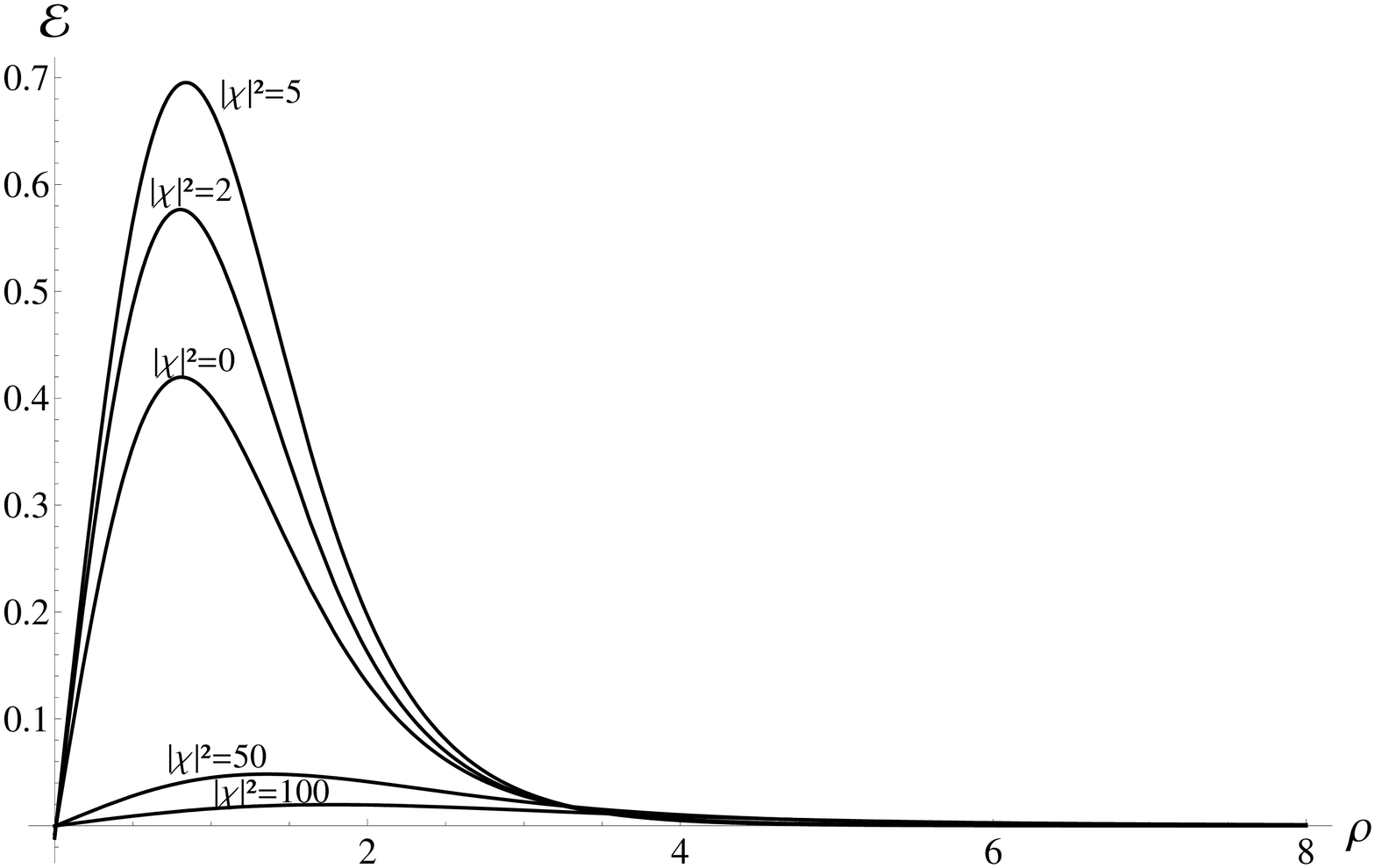}\label{fig:dbsemi}
\caption{The magnetic and electric field of an
$\left\{n=1,m_N/v^2=1,\kappa=100\right\}$ vortex with $|\chi|^2=
0,2,5,50,100$.} \label{dbsemi}
\end{figure}

Concerning the angular momentum for the semi-local vortices,
the component $T_{0\varphi}$ of the relevant energy-momentum
tensor component gets just an extra term $\Delta T_{0\varphi}$
with respect to the local case,
\begin{align}
\Delta T_{0\varphi} =& \,\frac{\eta_N^2}{e^2}\, \left( \sum_e
(D_0\xi_e)^*(D_\varphi \xi_e) +
\text{h.c.} \right) \nonumber\\
=& \mp\, 2 \frac{\eta_N}{e^2}\,h_n  (n-a_N) \, \eta_N^2\, |\xi_e|^2
\end{align}
and then
\begin{align}
T_{0 \varphi} &=\pm\frac{2}{e^2}\, h'_N\, a'_N \pm 2\, \eta_N^2\,
h_N\, \left(q_N^2 + |\xi_e|^2\right)\,a_N
\mp 2\,n\, \eta_N^2\, h_N\,  |\xi_e|^2
\end{align}
Using Gauss' Law
\begin{equation}
\rho^{-1} \, \frac{ d }{d\rho} h_N(\rho)\, \rho) = \eta_N^2\, h_n
\left(q_N^2+|\xi_e|^2\right) \mp \frac{\kappa}{4\pi} \rho^{-1}\,
a'_N
\end{equation}
we can write
\begin{equation}
T_{0\varphi} = \pm \frac{2}{e^2}\frac{1}{\rho} \frac{d}{d\rho}
(h'_N\, a_N\, \rho) + \frac{\kappa}{4\pi}\, \frac{1}{\rho}
\frac{d}{d\rho} (a_N^2) \mp 2\, n\,\eta_N^2\, h_N\,  |\xi_e|^2
\end{equation}
where the two first term coincide with the local vortex value for $T_{0\varphi}$.
Taking into account  boundary conditions one now has
\begin{equation}
J = 2\pi \int    T_{0\,\varphi} \,\rho\, d\rho = -\frac{\kappa}{2}\,
n^2 \pm 2\, n\, \eta_N^2\, \int h_N\, |\xi_e|^2\, \rho d\rho
\end{equation}
Thus, unlike the case of local vortex, the angular momentum is
not simply quantized in terms of $\kappa n^2$ but  it depends
explicitly on the size of the semi-local vortex.

\section{Discussion}

A rich spectrum of  non-Abelian vortices in theories where the
dynamics of the gauge field is governed both by a Yang-Mills
and a Chern-Simons action was shown to exist in
refs.\cite{CollieTong}-\cite{Collie}, where the low-energy
vortex dynamics was described in terms of a gauged sigma model
on the vortex worldline. Although the BPS equations were
obtained, explicit solutions were not presented and this was
precisely the main objective of our work. To this end,
 we proposed  an axially symmetric ansatz leading to BPS
vortex solutions for a Yang-Mills-Chern-Simons $U(N)$ gauge
theory coupled to scalars  when the number of flavors $N_f \geq
N$, analyzing the electric and magnetic properties of the local
($N_f = N$) and semi-local ($N_f>N$) vortices.

A first interesting feature of the local vortex solutions
concerns the localization of the magnetic and electric fields.
As expected, as the Chern-Simons coefficient $\kappa$ grows,
the maximum of  magnetic moves away from the vortex center and
the electric field, also with an annulus shape, starts to
develop. Semi-local vortices exhibit a similar behavior except
that   ${\cal B}$ and  ${\cal E}$ have a long range power
falloff instead of an exponential one (only the  real scalar
field keeps its exponential falloff behavior). Another
difference between local and semilocal vortices concerns the
angular momentum which is a purely topological object in the
former case while it depends on the vortex size in the later
semilocal case

Our investigation started from the Lagrangian proposed in
\cite{Collie} with the scalar potential and constants chosen so
as to guarantee the possibility of an ${\cal N} = 2$
supersymmetric extension, this ensuring the existence of BPS
equations \cite{T}-\cite{SY}. Actually, it  would be of
interest to investigate the properties of the supersymmetric
model and to construct the low-energy effective action
describing moduli dynamics and analyze its properties both at
the classical and quantum level, following the approach
presented in ref.\cite{Aldrovandi:2007nb} for the pure
Chern-Simons case. We hope to discuss this issues in a future
work.

%%%%%%%%%%%%%%%%%%%%%%%%%%%%%%%%%%%%%%%%%%%%%%%%%%%%%%%%%%%%%%%%%%%%%%%%%%%%%
%%%%%%%%%%%%%%%%%%%%%%%%%%%%%%%%%%%%%%%%%%%%%%%%%%%%%%%%%%%%%%%%%%%%%%%%%%%%%
%%%%%%%%%%%%%%%%%%%%%%%%%%%%%%%%%%%%%%%%%%%%%%%%%%%%%%%%%%%%%%%%%%%%%%%%%%%%%

\begin{acknowledgements}
This work was partially supported by
UNLP, UBA, CICBA, CONICET and ANPCYT and MinCyT. One of us
(FAS) whihes to thank the ECM Department of the Universitat de
Barcelona for hospitality during completion of this work.
\end{acknowledgements}

%%%%%%%%%%%%%%%%%%%%%%%%%%%%%%%%%%%%%%%%%%%%%%%%%%%%%%%%%%%%%%%%%%%%%%%%%%%%%
%%%%%%%%%%%%%%%%%%%%%%%%%%%%%%%%%%%%%%%%%%%%%%%%%%%%%%%%%%%%%%%%%%%%%%%%%%%%%
%%%%%%%%%%%%%%%%%%%%%%%%%%%%%%%%%%%%%%%%%%%%%%%%%%%%%%%%%%%%%%%%%%%%%%%%%%%%%

%%%%%%%%%%%%%%%%%%%%%%%%%%%%%%%%%%%%%%%%%%%%%%%%%%%%%%%%%%%%%%%%%%%%%%%%%%%%%
%%%%%%%%%%%%%%%%%%%%%%%%%%%%%%%%%%%%%%%%%%%%%%%%%%%%%%%%%%%%%%%%%%%%%%%%%%%%%

\end{document}